\documentclass[11pt]{article}
\usepackage{aas_macros,graphicx,deluxetable,natbib}
\def\Msun{M$_{\odot}$}
\def\Msunyr{M$_{\odot}$ yr$^{-1}$}

\begin{document}

\begin{center}
{\huge \bf Radio Supernovae: \\
Circum-Stellar Investigation (C.S.I.) of Supernova Progenitor Stars}

\vspace{8 mm}

{Christopher J. Stockdale \\
Marquette University \\ 
Physics Department \\
PO Box 1881 \\
Milwaukee, WI, 53214-1881 \\
christopher.stockdale@mu.edu}

\vspace{5 mm}

{Kurt W. Weiler \\
Naval Research Laboratory}

\vspace{5 mm}

{Nino Panagia \\
Space Telescope Science Institute}

\vspace{5 mm}

{Richard A. Sramek \\
National Radio Astronomy Observatory}

\vspace{5 mm}

{Schuyler D. Van Dyk \\
Spitzer Science Center}

\vspace{5 mm}

{Stefan Immler \\
Goddard Space Flight Center, NASA}

\vspace{5 mm}

{Dave Pooley \\
University of Wisconsin}

\vspace{5 mm}

{J. M. Marcaide \\
Universitat de Val\`{e}ncia}

\vspace{5 mm}

{S. Ryder \\ 
Anglo-Australian Observatory}

\vspace{5 mm}

{Matthew T. Kelley \\
University of Nevada Las Vegas}

\vspace{5 mm}

{Christopher L.~Williams \\
Massachusetts Institute of Technology}

\end{center}

\newpage 

\section{Why are Radio Observations of Supernovae Important?}

Supernovae (SNe) are among the most energetic phenomena in the Universe, 
releasing $\ge10^{51}$ erg in a few seconds with shock speeds ranging from a few 1,000 km s$^{-1}$ to $\sim230,000$ km s$^{-1}$ (for the 
extreme Type Ic SN 1998bw). 
They are integrally linked to many phenomena of importance to modern 
astrophysics  -- formation of black holes and neutron stars, at least 
some types of gamma-ray bursters, powering starburst galaxies, 
nucleosynthesis, cosmology probes, and distribution of heavy elements 
and energy into the interstellar medium (ISM).  
 
SNe come in two basic types, type Ia (thermonuclear detonation of white dwarfs) 
and types Ib/Ic/II (resulting from the core-collapse of massive stars).  While no radio emission has ever been detected from type Ia SNe (Panagia et al. 2006), 75\% of detected radio SNe are type II and 25\% are type Ib/Ic.  Type II SNe are thought to be the deaths of stars between 8\Msun \ and 40\Msun , while type Ib/Ic SNe generally thought to be exploding Wolf-Rayet stars (M$>40$\Msun; Conti et al. 1987; Humphries et al. 1985) or lower mass stars in interacting-binary systems (Uomoto 1986; Podsiadlowski et al. 1992).

In the $\sim$10,000 years prior to explosion, 
the progenitor star is believed to undergo a period of enhanced mass-loss creating a dense region
of circumstellar material (CSM) which is propagating outward at $\sim10$ km s$^{-1}$ (Chevalier 1982a).  
The UV flash from the SN explosion ionizes the CSM through which the SN blastwave propagates and
heats the CSM to T$_{\rm CSM} \sim 10^5$ K. In the early phase after the SN explosion, the average blastwave speed 
 is on the order of 
$\sim10^4~{\rm km~s}^{-1}$.  In the forward shock created at the interface
where the blastwave overruns the CSM, radio emission 
is generated by relativistic electrons swept up in the intense magnetic 
fields associated with the blastwave (Chevalier 1982a; Weiler et al. 1986, 2002, 2005).  
A reverse shock is also created with a typical ``inward'' speed of
 $\sim$10$^3~{\rm km~s}^{-1}$ relative to the forward shock.  
The reverse shock propagates back into the rapidly outward-moving
SN ejecta, resulting in X-Ray emission.  A detailed discussion of X-Ray Supernovae has been submitted in a white paper prepared by Dave Pooley. 

\begin{figure}[t]
\begin{center}
\hspace{1cm}
\rotatebox{-90}{\includegraphics[width=.50 \textwidth]{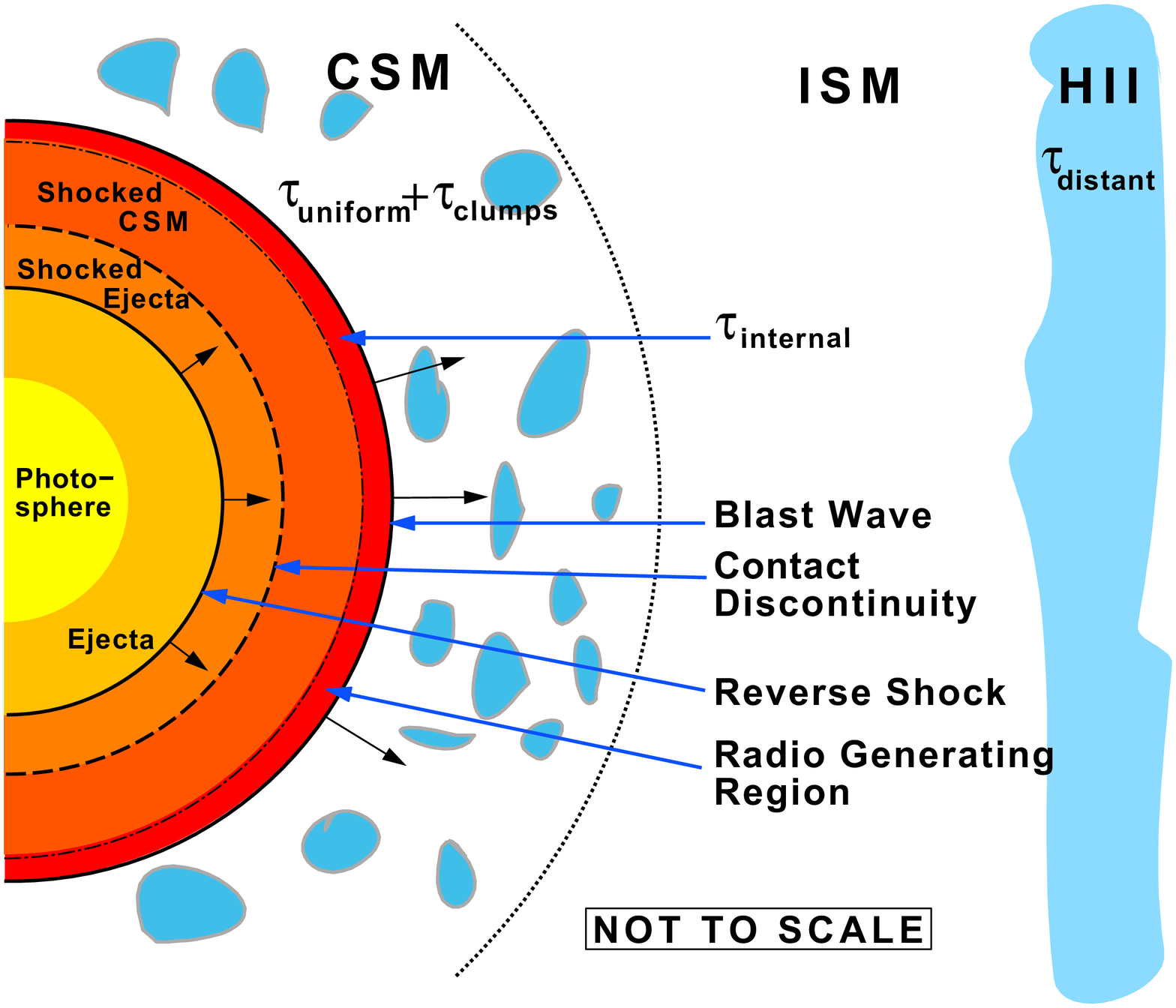}}
\end{center}
\caption{}{\it
Cartoon image of the SN blastwave interacting with the CSM (Stockdale et al. 2007).  
The radio emission originates just behind the blastwave and the X-ray emission is 
dominated by emission from the shocked CSM and ejecta, heated by the reverse shock.}
\label{cartoon}
\end{figure}

Progenitor stars have
been identified in only a few cases, and none have been the subject of extensive monitoring prior to 
explosion (Li et al. 2007).  Further, optical observations only sample the SN photosphere 
and ejecta at early times after the explosion.  Figure~\ref{cartoon} identifies significant regions  to the emission of radio
and X-rays associated with the SN blastwave interacting with the CSM.  

The modeling of the SN blastwave interaction with the CSM
allows the determination of a number of physical properties of 
SNe from radio observations.  One of these 
is the mass-loss rate from the SN progenitor prior to explosion. 
From the Chevalier (1982a,b) model, the turn-on of the radio emission 
for RSNe provides a measure of the presupernova mass-loss rate to wind 
velocity ratio ($\dot M/v_{\rm w}$).  
Weiler et al. (1986) derived this ratio for the case of pure, external absorption by a homogeneous medium. 
There are several possible physical mechanisms responsible for the absorption of the radio emission by the CSM, characterized the the effective optical depth at a given radio frequency  $\tau_{\rm eff}$.  The generalized equation (16) of
Weiler et al. (1986)  relates the mass-loss rate at given epoch to the radio determined optical depth as

\begin{eqnarray}
\label{eq11}
\frac{\dot M (M_\odot\ {\rm yr}^{-1})}{( v_{\rm w} / 10\ {\rm km\ s}^{-1} )} & = &
3.0 \times 10^{-6}\ <\tau_{{\rm eff}}^{0.5}> \ m^{-1.5} {\left(\frac{v_{\rm s}}{10^{4}\ {\rm km\ s}^{-1}}\right)}^{1.5} \times \nonumber \\ & & {\left(\frac{t_{\rm i}}{45\ {\rm days}}\right) }^{1.5} {\left(\frac{t}{t_{\rm i}}\right) }^{1.5 m}{\left(\frac{T}{10^{4}\ {\rm K}} \right)}^{0.68} .
\end{eqnarray}

Weiler et al. (2002) and Sramek et al. (2003) discuss in the detail how radio
light curves of supernova can be fit to a parameterized model accounting 
for the nature of the CSM.  Weiler et al. (2002) 
were able to identify at
least three possible absorption regimes: 1) absorption by a homogeneous
external medium, 2) absorption by a clumpy or filamentary external 
medium with a statistically large number of clumps, and 3) absorption by
a clumpy or filamentary  medium with a statistically small number of
clumps.  Each of the three cases requires a different formulation for
$<\tau_{{\rm eff}}^{0.5}>$.  Only with a well-sampled radio light curve is 
it possible to robustly model the CSM and determine the appropriate mass-loss
history for a given SNe.

Radio observations probe the SN blastwave's interaction with the CSM
 and the absorbing properties of the CSM in the radio emitting region and 
in the region along the line of sight to the observer.  Using the radio observations, it is possible to map the circumstellar density 
profiles and to measure and
compare mass-loss rates of the SN progenitor stars.  Studying the radio emission from core-collapse SNe,  astronomers can explore these three key questions: 
\begin {enumerate}
\item How do the massive progenitor stars of core-collapse SNe evolve in the thousands of years prior to their explosion?
\item What are the physical processes of the absorbing medium of the early radio emission?  
\item Do these SNe simply transition smoothly into Supernova Remants (SNRs) or is there a fading as they overrun their CSM and later re-brightening as the blastwave begins to encounter the ISM?    
\end{enumerate}

These questions can be explored through an active program that searches for radio emission from nearby core-collapse SNe and the subsequent long-term monitoring of radio SNe.  
The early observations, those made within 1-2 months of explosion, probe the mass-loss history of the progenitor in the hundreds
of years immediately preceding the SN explosion.  Later observations probe the evolution of the CSM at progressively earlier epochs of the progenitor star's pre-explosion evolution.  The multi-frequency radio light curve 
transitions from optically thick to optically thin are an important probe 
of the absorption physics of the CSM (Chevalier 1982; Weiler et al. 1986, 2002, 2005).  The key observational objectives of an effective radio SN program include:
\begin{itemize}
\item an aggressive, but selective Target-of-Opportunity Program); and
\item detect and chart the rapid evolution of radio emission from new core-collapse SNe within 25 Mpc (weekly to monthly); and
\item monitor the radio evolution of SNe within the last 20 yrs (semi-annually).
\end{itemize}

\begin{figure}[t]
\begin{center}
\includegraphics[width=0.5\textwidth]{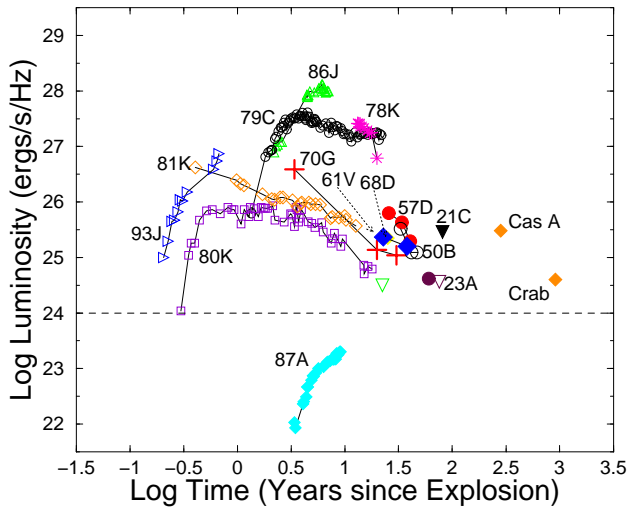}
\end{center}
\caption{\label{lc} 
{\it L-band (20 cm) radio light curves for a number of historical radio SNe, plotting the log of the spectral luminosity vs. the log of the
time since explosion.
The dashed line indicates a typical sensitivity of 10$^{24}$ erg s$^{-1}$ Hz$^{-1}$, roughly 12 hours of VLA observation for a source at $\sim$5 Mpc.    There is a 300 year gap between SN 1923A, the oldest radio SN, and the youngest SNRs!  See Stockdale 
et al. 2006 and references therein for the data regarding each SN.}}
\end{figure}

Because only a few SNe are 
detectable  at reasonable observing times ($<1$ hr per frequency measurement) using the 
Very Large Array (VLA)\footnote{The VLA telescope of the National 
Radio Astronomy Observatory (NRAO) is operated by Associated
Universities, Inc. under a cooperative agreement with the National Science 
Foundation.}, 
a carefully-planned and aggressive Target-of-Opportunity program is essential for identifying which SNe
are candidates for further radio study.  This complements the Long Term Monitoring 
which follows the evolution of the radio emission for as long as 75 years after explosion, as is the case of SN~1923A in
the spiral galaxy M83 (Stockdale et al. 2006; Eck et al. 1998).  Prior to SN~1923A, the last observed SNe with detected radio emission were Cas~A 
(dated to the late 17th century) and SN~1604 (observed by Kepler).  Due to the scarcity of observational evidence, the evolution of SNe into supernova remnants (SNRs) is poorly understood.  Because the non-thermal radio emission from the blastwave/CSM interaction persists for decades, radio
observations are the only feasible method to study this transition phase.  Figure~\ref{lc} illustrates the lack of measurements of 100-300 year old SNe probing this transition Stockdale et al. 2006).

{\bf Together, Radio and X-Ray measurements provide a deeper insight in the pre-explosion evolution of core-collapse supernova progenitors.}

\section{The Current State of Radio Supernovae Studies}

At present, there are on-going RSN observing programs
on six continents at a variety of facilities, including the 
Giant Metrewave Radio Telescope (India) and the Australia Telescope Compact Array, but the VLA remains the most sensitive and useful instrument for the detection and monitoring
of these transient sources.   The VLA has been monitoring SNe for more than 25 years initiated by K. W. Weiler, N. Panagia, and R. A. Sramek.  The VLA observing programs have detected dozens of new radio SNe.  After discovery, the radio emission is monitored at multiple frequencies until the emission has completed faded under a Long Term Monitoring Program.  This may take anywhere from months to decades, depending on the nature
of the progenitor.  Because many of these sources evolve over such long time-scales, a complete analysis of the radio
evolution often requires years of observations to yield significant results.  

\subsection{Target of Opportunity Program} 

\begin{figure}[t]
\begin{center}
%\hspace{1cm}
\includegraphics[width=0.5\textwidth]{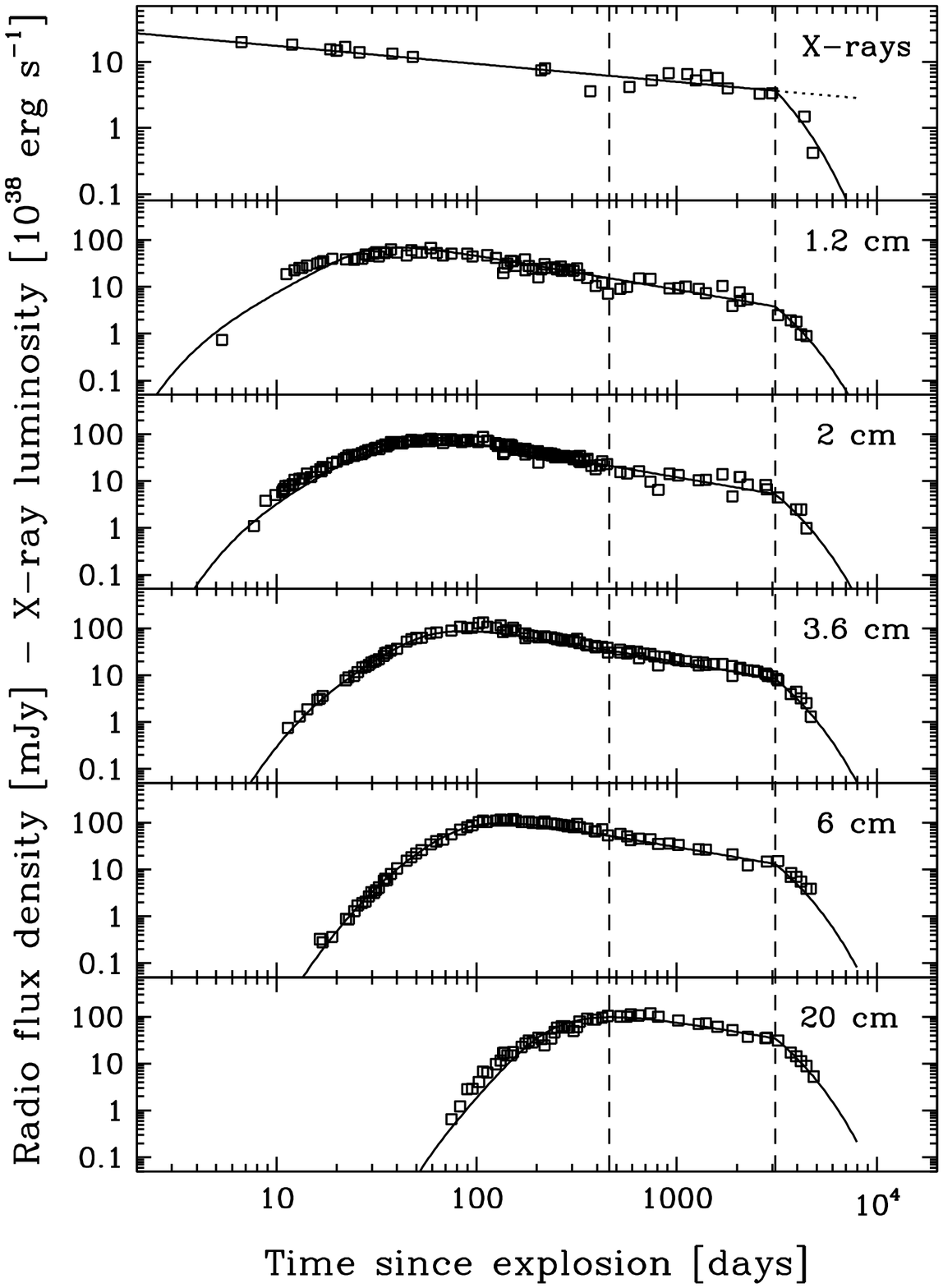}
\end{center}
\caption{}{{\it SN~1993J is the most well studied radio SN outside of the Local Group, with over 650 radio observations taken mostly with
the VLA, the Ryle telescope, and the Giant Meter-wave Radio Telescope (GRMT).  The wavelengths for each light curve are shown in the figure.  There is a slight increase in flux density as compared to the model prediction at 20 cm following the turn-over, also present in the other bands.  There is a clear correlation between the X-ray and radio light curve evolution.  Also, note the achromatic drop at day $\sim~3,100$  (Weiler et al., 2007).}}
\label{93J}
\end{figure}

In the last eight years, seven new type II radio SNe have been discovered, and there have also been more than a dozen unsuccessful searches.  An excellent example of the need for early detection and frequent radio observations of new SNe is SN~1993J.  
As shown in Figure~\ref{93J}, SN~1993J  represents the most complete set of radio observations from its initial radio detection to the sudden decline as the CSM becomes optically thin.  For SN~1993J, Weiler et al. (2007)  have been able to determine that both synchrotron self-absorption and free-free absorption by the CSM contribute to the radio optical opacity.  This determination was only possible because of early and daily monitoring ($<10$ days after explosion). 

\subsection{Long Term Monitoring of Monitoring of Recent Supernovae}

Following the their initial detections, these SNe, which span a variety of optical classifications, are now the subject of on-going VLA observations.  SNe~2001gd, 2004et, and 1986J  are three SNe that are currently being monitored by the VLA. 

\begin{figure}[t]
  \centering
  \begin{minipage}[c]{8cm}
   {\includegraphics[width=0.8\textwidth]{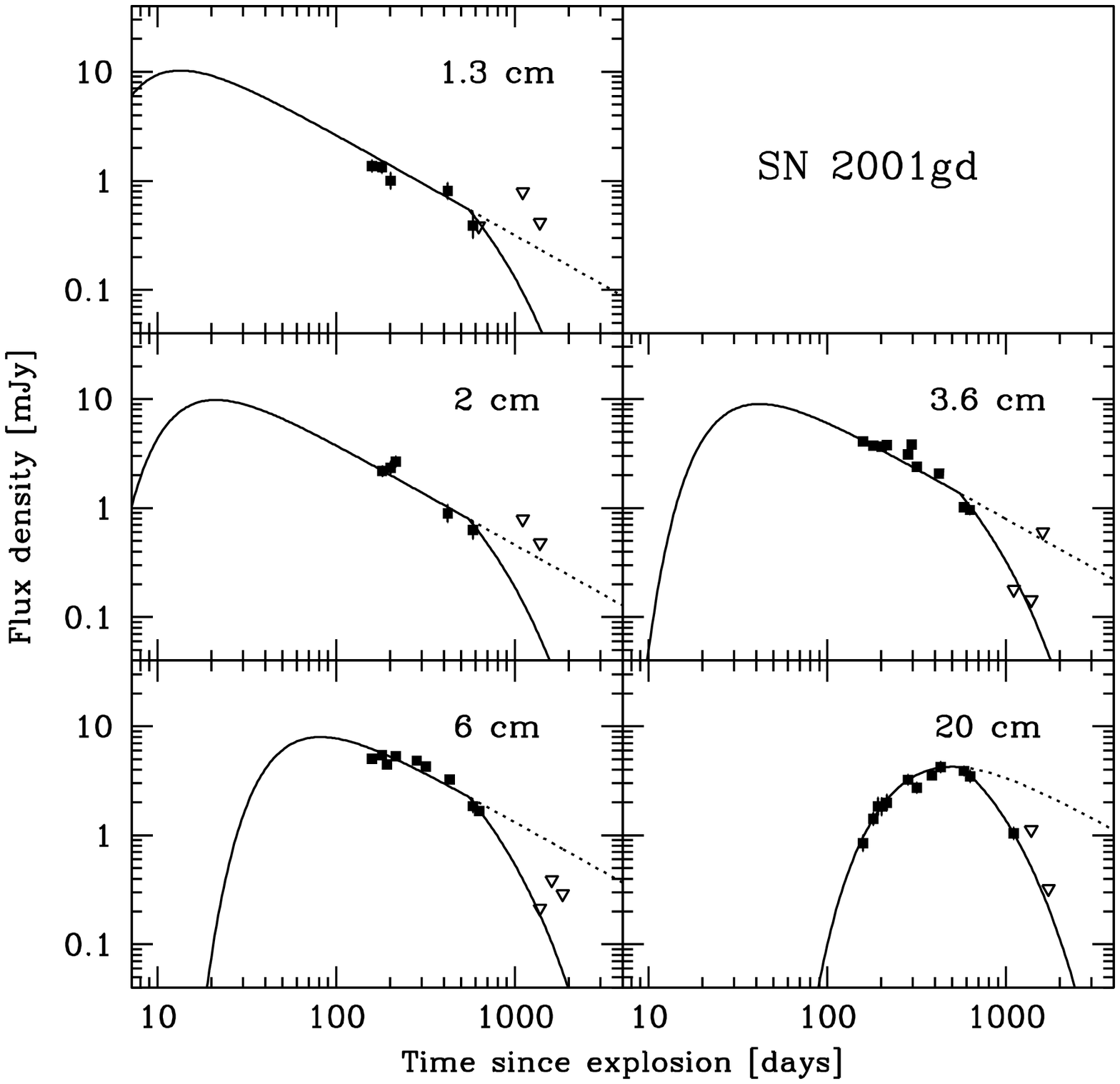}}  
  \end{minipage}
  \begin{minipage}[c]{8cm}
    \rotatebox{-90}{\includegraphics[width=0.8\textwidth]{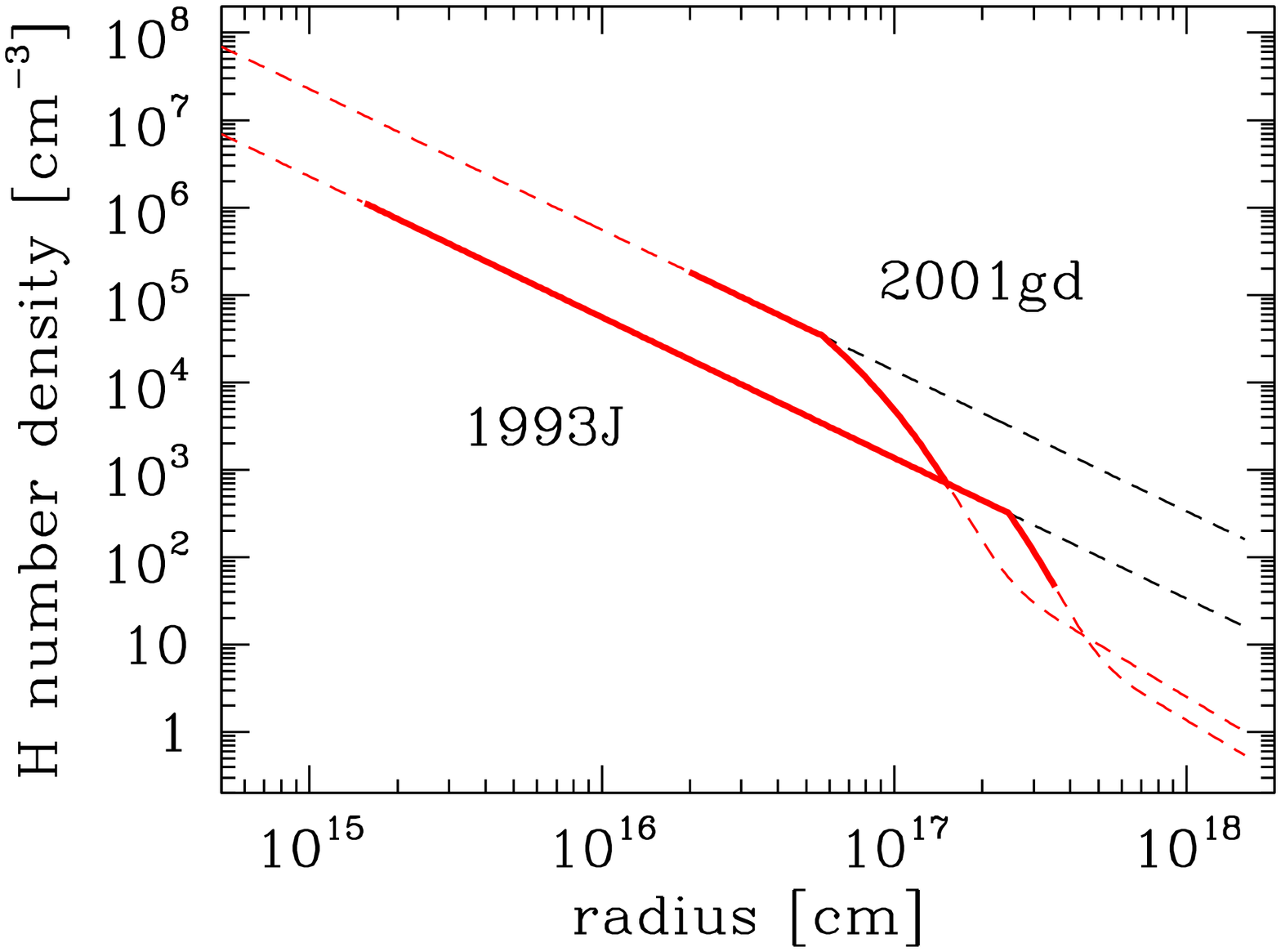}}    
  \end{minipage}
  \caption{\it  Left, the VLA light curves of SN~2001gd in the five main VLA bands.  Note the achromatic break at day $\sim550$, similar to that SN~1993J.  Right, the ionized hydrogen density plots as a function of radius for SNe~2001gd and 1993J.  (Stockdale et al. 2007; Weiler et al. 2007)}
  \label{01gdLC}
\end{figure}

{\bf SN~2001gd:}  Stockdale et al. (2007) indicate that the evolution of SN~2001gd [a type IIb SN] diverges from the simple Chevalier (1982a) model of constant, wind-driven CSM.  The radio light curves of SN~2001gd, shown in Figure~\ref{01gdLC}, suggest a similar evolution to that of SN~1993J (see Figure~\ref{93J}),
 typified by a dramatic decline in their radio emission at all 
frequencies in VLA observations near day $\sim 550$ (Stockdale et al. 2007).  SNe~1993J and 2001gd   are both type IIb SNe, and their declines indicate abrupt increases in mass-loss just prior to the explosion that stripped the hydrogen shells from the progenitors (Stockdale et al. 2007; Weiler et al. 2007).   In Figure~\ref{01gdLC}, the radial density of the CSM (assuming a typical 10 km s${-1}$ pre-explosion wind velocity) for SN~1993J  and  SN~2001gd is presented.  For both SNe, Stockdale et al. (2007) and Weiler et al. (2007) interpret the abrupt change in the radio evolution as indication of a substantially enhanced wind developing no earlier than 8,000 years prior to the explosions of the progenitor stars.  Future VLA observations will continue to probe this exponential decline.

{\bf SN~2004et:} SN~2004et  [a type II-P SN] was thought to be a fairly intact hydrogen envelope with a relatively low mass progenitor star.  Chugai et al. (2007) predicted the radio evolution of SN~2004et and other type IIP SNe should follow the Chelavier (1982) model.  Kelley et al. (2009) reported the recent observations of SN~2004et that indicate an abrupt change in the progenitor mass-loss rate, the nature of which is yet to be determined.  Future VLA observations will provide critical confirmation as to whether 
({\bf 1.}) this decline continues to persist indicating an abrupt cessation to the CSM which may be an indication of significant change in the mass-loss history of the progenitor stars or 
({\bf 2.}) this is a fluctuation in the CSM due to a previously undetected binary companion akin to the behavior of SNe~2001ig (Ryder et al. 2004, 2005) and 1979C (Montes et al. 1998) .  From these preliminary measurements, Kelley et al. (2009) estimated the progenitor mass loss rates 
to the order of $10^{-5}$ \Msunyr for both SNe, assuming a constant velocity, filamentary medium as described in
Weiler et al. (2002).  

{\bf SN~1986J:}  Recent Very Long Baseline Interferometry (VLBI) observations of SN~1986J have yielded new insight into the study of core-collapse SNe.  Bartel et al. (2007) have completed a series of high-resolution radio imaging of this type IIn SN, the first ever detected in the radio, prior to an optical detection.  Bartel et al. (2007) indicate the likely formation of a neutron star, the first historical SN in the last 100 years to make this transition.   The initial analysis of the recent VLA data appears to support this discovery, reporting an inversion of the spectral index at higher frequencies.  A thorough analysis of the VLA observations is warranted to determine how the VLA observations can be used to distinguish between radio emission 
from blastwave/CSM interactions and the formation of pulsar wind-nebula in other SN observations where VLBI measurements are not feasible.

\section{The Future of Radio Supernovae}

Current observing programs with the VLA  typically yield  1-2 new radio detections per year.  Because past observations have shown that astronomers only detect in the radio a small percentage of newly discovered SNe (currently $<5\%$), only those objects which have a peak optical m$_v < 13$ or
a distance  $<25$ Mpc have been observed to maximize our use of the VLA observing allocation.  With the soon to be commissioned Expanded VLA (EVLA), we will be able to expand our investigation to include SNe out to 75 Mpc, which would increase our detection of new radio SNe by more than an order of magnitude.  This will allow astronomers to explore the evolution of a broad variety of SNe for a longer time.  We will be able to discern the comprehensive mass-loss histories of the SN progenitor stars, as done for SN 1993J, for at least one SN per year, rather than one per decade.  There are only one, or at best two, well sampled radio SNe of any given optical classification.  This is insufficient to make any true determinations concerning the CSM structure for an entire optical classification.  It is essential that we study all core-collapse SNe as a unified class, as the zoology of optical classifications (Ib, Ic, IIP, IIn, IIL, IIb...) has already begun to blur with SNe evolving from one optical type into another.

{\bf With the EVLA, astronomers will study of a broad variety of types of core-collapse SNe, observationally determining the detailed mass-loss rates that are essential in discriminating models for progenitor evolution.}  A database of a variety of SN types and their early pre-explosion evolutionary stages will be built.  The research performed with the EVLA will consist of 
the most comprehensive studies available on the evolution of the radio emission of supernovae.  This will allow a unique insight into
the evolution of high-mass, SN-progenitor stars. {\bf With the EVLA, astronomers will be able to probe the eventual transition of SNe from explosion into SNRs and  will be to study the galactic environments
in which SNe are formed.}

{\bf With the Square Kilometer Array (SKA), normal radio SNe would be detectable to $\sim200$ Mpc.}  It may even become possible to detect radio emission from type Ia SNe, or at least to establish more stringent constraints on wind-driven, mass-loss rates of the white dwarfs' companion stars.

\section{References}

\noindent
Bartel et al. 2007, ApJ,  668, 2, 924

\noindent
Chevalier 1982a, ApJ, 258, 302

\noindent 
Chevalier 1982b, ApJ, 259, L85

 \noindent
Chevalier \& Fransson,  2006, ApJ, 651, 1, 381

 \noindent
Chugai et al.  ApJ, 662, 2, 1136

 \noindent
Conti et al. 1983, ApJ, 274, 302

 \noindent
Eck et al. 1998, ApJ, 508, 664

 \noindent
Filippenko 1997, ARAA, 35, 309

 \noindent
Humphreys, Nichols, \& Massey 1985, AJ, 90, 101

 \noindent
Kelley et al. 2009, in prep.

 \noindent
Li et al.  2007, ApJ, 661, 2, 1013

 \noindent
Montes et al. 1998, ApJ, 532, 2, 1124

 \noindent
Panagia et al. 2006, ApJ, 646, 369

 \noindent
Podsiadlowski, Joss, \& Hsu 1992, ApJ, 391, 246

 \noindent
Ryder et al. 2004, MNRAS, 349, 1093

 \noindent
Ryder et al. 2005, IAU Colloquium, 192, 123

 \noindent
Sramek \& Weiler 2003, Lecture Notes in Physics, 598, 145

 \noindent
Stockdale et al. 2006, AJ, 131, 2, 889

 \noindent
Stockdale et al.  2007, ApJ, 671,1, 689

 \noindent
Uomoto 1986, ApJ, 310, L35

 \noindent
Weiler et al. 2002, ARAA, 40, 387-438

 \noindent
Weiler et al. 1986, ApJ, 301, 790

\noindent
Weiler et al. 2005, ASP Conference Series 342, 290

\noindent
Weiler, et al. 2007, ApJ, 671, 2, 1959

\end{document}